\documentclass[english,12pt]{iopart}

\usepackage[T1]{fontenc}
\usepackage[latin1]{inputenc}
\usepackage{amssymb}
\usepackage{graphicx}

\makeatletter

\@ifundefined{textcolor}{}
{%
 \definecolor{BLACK}{gray}{0}
 \definecolor{WHITE}{gray}{1}
 \definecolor{RED}{rgb}{1,0,0}
 \definecolor{GREEN}{rgb}{0,1,0}
 \definecolor{BLUE}{rgb}{0,0,1}
 \definecolor{CYAN}{cmyk}{1,0,0,0}
 \definecolor{MAGENTA}{cmyk}{0,1,0,0}
 \definecolor{YELLOW}{cmyk}{0,0,1,0}
 }

\usepackage{epstopdf}

\usepackage{color}

\makeatletter


\def\c{s}
\def\d{s}
\definecolor{BrickRed}{cmyk}{0,0.89,0.94,0.28}%%%PANTONE 1805
\definecolor{MidnightBlue}{cmyk}{0.98,0.13,0,0.43}%%%PANTONE 302
\definecolor{DarkGreen}{rgb}{0,0.7,0.1}

\newcommand{\comm}[1]{\if\c\d{{\color{MidnightBlue}\{\small \sc #1\}}}\else{}\fi}
\newcommand{\add}[1]{\if\c\d{{\color{magenta}#1}}\else{#1}\fi}
\newcommand{\drop}[1]{\if\c\d{{\color{DarkGreen}[[#1]]}}\else{}\fi}

\makeatother
\makeatother

\makeatother

\usepackage{babel}

\begin{document}

\title{One-dimensional array of ion chains coupled to an optical cavity }

\author{Marko Cetina\footnote{currently at IQOQI, Innsbruck, Austria}, Alexei Bylinskii*, Leon Karpa, Dorian Gangloff, Kristin M. Beck, Yufei Ge, Matthias Scholz\footnote{currently at TOPTICA Photonics AG, R\&D Division, Lochhamer Schlag 19, 82166 Gräfelfing, Germany}, Andrew T. Grier\footnote{currently at \'Ecole Normale Sup\'erieure, Paris, France}, Isaac Chuang, and Vladan Vuleti\'{c}}

\address{Department of Physics, MIT-Harvard Center for Ultracold Atoms, and
Research Laboratory of Electronics, Massachusetts Institute of Technology,
Cambridge, Massachusetts 02139, USA}

\ead{* abyl@mit.edu}

\date{\today}
\begin{abstract}
We present a novel hybrid system where an optical cavity is integrated
with a microfabricated planar-electrode ion trap. The trap electrodes produce
a tunable periodic potential allowing the trapping of up to 50 separate ion
chains spaced by 160 $\mu$m along the cavity axis. Each chain can contain
up to 20 individually addressable Yb\textsuperscript{+} ions coupled
to the cavity mode. We demonstrate deterministic distribution of ions
between the sites of the electrostatic periodic potential and control
of the ion-cavity coupling. The measured strength of this coupling should allow access to the strong collective coupling regime
with $\lesssim$10 ions. The optical cavity could serve as a quantum
information bus between ions or  be used to generate a strong wavelength-scale
periodic optical potential. 
\end{abstract}


\submitto{\NJP}

\maketitle

\section{Introduction}

The coupling of trapped atomic ions to optical cavities constitutes
a promising route for scaling quantum information processing (QIP)
to larger ion numbers and exploring a range of possibilities in quantum
simulation (QSim). Ions are an ideal building block for both QIP \cite{Blatt2008,Haffner2008}
and QSim \cite{Johanning2009a,Blatt2012,Schneider2012a} thanks to
exquisite individual control and strong Coulomb interactions. These properties
have enabled quantum gate fidelities close to the threshold where error
correction protocols could guarantee fault-tolerant computation \cite{Benhelm08}.
Entanglement of up to 14 ions has been achieved \cite{Monz2011a} and
small spin networks of up to 9 spins have been simulated \cite{Friedenauer2008,Kim2010a,Islam2011,Kim2011}. All the schemes that have achieved these milestones involve coupling the ions via common motional modes of 1D Coulomb crystals that are formed when ions are trapped in the same trap. Unfortunately, the increasing number of collective motional modes in longer ion chains makes it difficult to scale this approach to larger ion numbers. One way around these
limitations is to interface ions with a quantum bus in the form of ``transport'' ions, photons or phonons.

One possible realization of a quantum bus is to physically transport ions; this requires reconfigurable
multi-zone traps and the ability to deterministically split ion crystals
and move ions without decoherence of the internal states that are
used for storing the quantum information \cite{Kielpinski2002}. Quantum
teleportation has been achieved in such a system \cite{Barrett2004a},
and the full toolbox for QIP has been demonstrated \cite{Home2009},
including transport with very little motional heating \cite{Blakestad2011}.

A photon-based quantum bus between ions is attractive because optical photons
are robust against decoherence over long distances and can be analyzed
and counted with high efficiency. Free-space coupling between photons and ions is weak, but has potential in probabilistic schemes \cite{Duan2010, MonroeArxiv2012}. Two ions have been entangled via
a long-distance photon bus in a probabilistic scheme with ~$10^{-8}$
success probability \cite{Olmschenk2009,Maunz2009}. Scalable, deterministic photon-based schemes
require an efficient, coherent ion-photon interface that can be provided by an optical cavity. High cavity finesse and small optical mode volume are two ways to reach strong coupling between cavity photons and atoms or ions \cite{Thompson1992,Kuhn2002,Gehr2010}. Another approach is to couple cavity photons to an ensemble of many atoms or ions so as to reach the strong collective coupling regime \cite{Black2005,Herskind2009a}.

Coupling of a single ion to a cavity has been demonstrated on a strong
dipole transition \cite{Guthohrlein2001} and on a weak quadrupole
transition with the drawback of slow coupling, susceptible to technical
decoherence \cite{Mundt2002a}. A high-finesse cavity coupled to
the weak leg of the Raman $S-P-D$ transition in a Ca\textsuperscript{+} ion has
enabled motional-sideband-resolved Raman spectroscopy \cite{Russo2009a,Stute2012b}
and efficient single photon generation \cite{Keller2004,Barros2009,Stute2012b}.
This has led to fast, high-fidelity and tunable entanglement between
an ion and the polarization state of the output photon \cite{Stute2012a}.
Despite these advances, fast coherent coupling between single ions
and cavities remains a challenge. Short micro-cavities, which enable
the strong coupling regime for neutral atoms \cite{Kuhn2002,Gehr2010} are promising, but have so far been difficult to combine with ion traps because of light-induced charging of the dielectric mirrors resulting in strong time-varying forces on the ions \cite{Sterk2012}.

Strong collective coupling between a cavity and separate ensembles of neutral atoms in the cavity has been used to entangle these ensembles via the cavity mode \cite{Simon2007b}. However, the use of a cavity photon bus for entanglement of separate ion ensembles has not yet been demonstrated. Two natural benefits of doing this with ions are the elimination of motional decoherence due to Lamb-Dicke confinement by electric fields, and the ability to combine the cavity photon bus with the QIP tools utilizing the common motional modes of co-trapped ions \cite{Lamata2011a}. Strong collective coupling between a large 3D Coulomb crystal of ions and a cavity has been achieved \cite{Herskind2009a}, leading to the observation of cavity electromagnetically induced transparency
\cite{Albert2011}.  Single-ion addressability, motional mode control and ground-state cooling are all difficult to achieve in 3D Coulomb crystals, one of the challenges being strong trap-driven radio-frequency micromotion. These problems can be avoided in 1D ion chains, which would therefore yield a more promising system when coupled to an optical cavity.

Another possibility for realizing a quantum bus between ions is through the exchange of phonons between microtraps \cite{Cirac2000a,Chiaverini2008}. Since this type of coupling scales as $d^{-3}$, where $d$ is the separation between microtraps, the individual microtraps need to be closely spaced. Sufficiently small-scale traps are difficult to operate in view of thermally activated heating of ions in close proximity to trap electrodes that scales roughly as $d^{-4}$ \cite{Blatt2008}. Nonetheless, coherent exchange of phonons in pairs of microtraps separated by a few tens of $\mu$m has been demonstrated with single-phonon Rabi frequencies of a few kHz \cite{Brown2011,Harlander2011}. The coupling can be enhanced by introducing more ions in each microtrap, and the scaling with ion number has been experimentally shown to be faster than linear \cite{Harlander2011}.

An alternative way to realize small-scale microtraps is to trap ions in an optical lattice, where the length scale is that of an optical wavelength. Although optical trapping of ions is difficult, single ions have been trapped in an optical dipole trap for milliseconds \cite{Schneider2010},  and recently microsecond trapping times of single ions in optical lattices have also been demonstrated \cite{Linnet2012,Enderlein2012}. A major challenge here is for the optical dipole force from the lattice to overcome the large electric forces acting on the charge of the ion. An optical cavity is a convenient way to build up very high lattice intensities for this purpose.

Ions in optical lattices are also promising for quantum simulations in periodic potentials. Compared to neutral atoms interacting via short-range, next-neighbor interactions \cite{Trotzky2008,Simon2011},
ions in optical lattices would enable the study of systems with strong
long-range interactions. In one dimension, the interplay between Coulomb interactions in an ion crystal and an incommensurate periodic potential can lead to phase transitions that are of interest in both the classical and quantum
regimes in the context of the Frenkel-Kontorova model \cite{Garcia-Mata2006,Benassi2011,Pruttivarasin2011a}.
This model is closely related to energy transport in crystals and
friction, and is of considerable interest in several domains of physics.
Optical lattices can be extended to two and three dimensions, where
frustrated spin Hamiltonians \cite{Schmied2008} and synthetic gauge
fields \cite{Bermudez2011} could be simulated with ions. Imaging individual particles in lattice sites is a capability of great interest for QSim in periodic potentials, and recently single-site resolution has been reached for neutral atoms \cite{Bakr2009,Sherson2010}. Ions in optical lattices naturally separate by many lattice sites owing to their strong Coulomb repulsion. The resulting large separations and the ability to freeze ions in space using tight electrostatic potentials are promising for high-resolution imaging of ion positions in a lattice. Such imaging would be conducive to studying many of the mentioned models.

\begin{figure*}[t]
\includegraphics[scale=0.5]{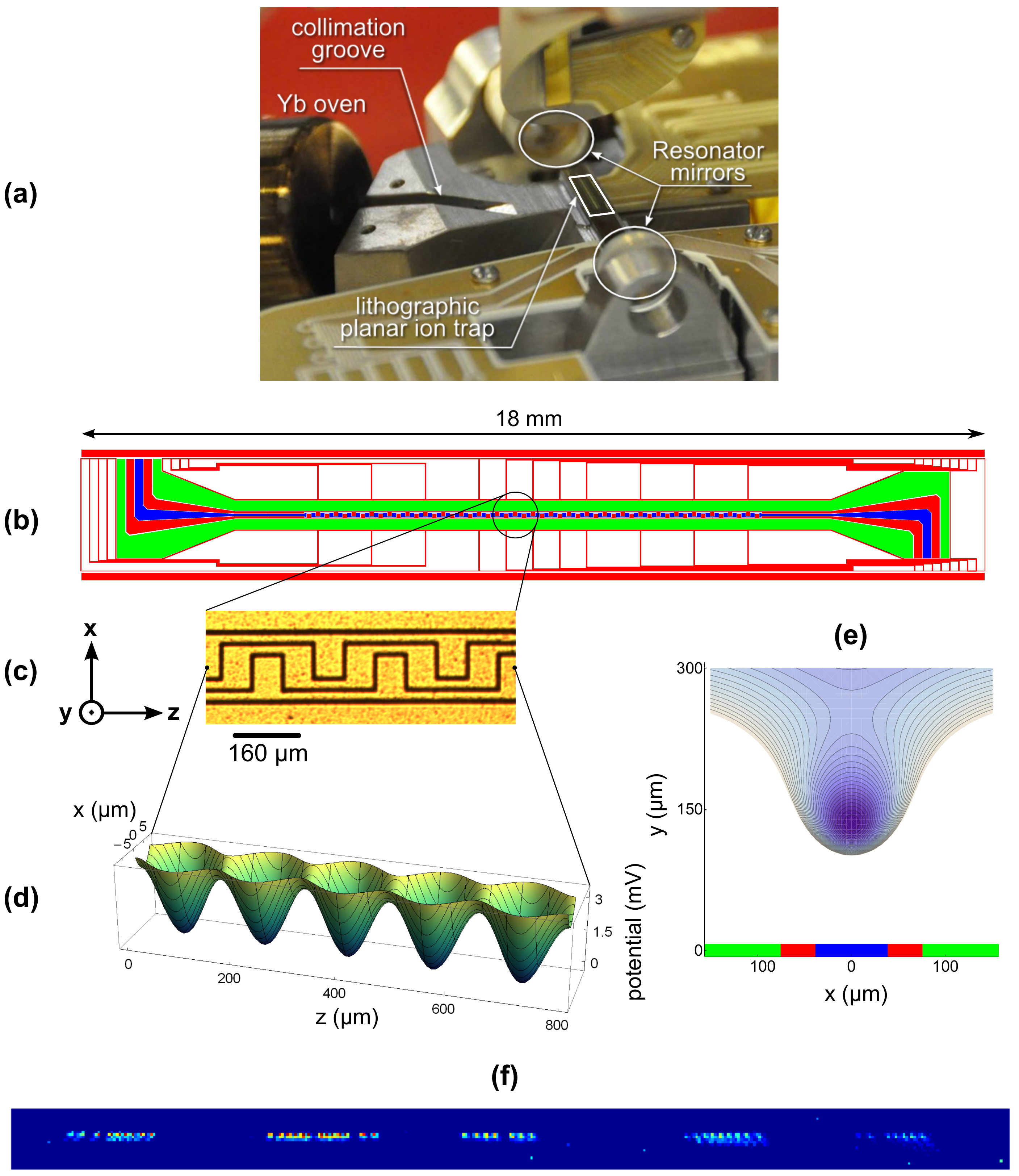}

\caption{\label{fig:trapfig1}
a) A picture of the experimental set-up, with the microfabricated planar electrode ion trap, the 2.2-cm-long optical cavity, and the Yb oven. 
b) Layout of the microfabricated trap chip. Outer RF electrodes are shown in
green, and the inner RF electrode is split into three periodic electrodes shown in blue and red. The 24 large rectangular DC electrodes are shown in white.
c) A portion of the inner periodic electrodes generating the electro-static periodic potential.
d) Periodic potential at the position of the ions 138 $\mu$m above the trap surface
for a -1 V DC voltage applied to the inner periodic electrode (blue) and +0.9 V to the outer periodic electrodes (red).
e) Pseudopotential produced in the radial directions by the RF voltage applied to the outer RF electrodes. The equipotential contours are spaced by 5 mV.
f) Image of 5 array sites containing small ion
chains where single ions can be resolved. The ions are illuminated
by light in the cavity mode, which is overlapped with the array. }
\end{figure*}

We present a novel system where all three
possibilities for a quantum bus (ionic, photonic and phononic) as well as the behavior of ions in periodic potentials can be explored. This is enabled by tunable periodic electrostatic and optical potentials and strong collective coupling of ions to an optical cavity. To our knowledge, this is the first system to integrate an optical cavity with a microfabricated planar electrode ion trap, producing a very versatile platform for QIP and QSim with ions and photons.

Our microfabricated planar electrode ion trap features a linear Paul
trap with a novel design of the inner electrode, which allows the Paul trap to be split into a periodic array of 50 separate traps by a tunable electrostatic potential. The sites of the array are separated by
160 $\mu$m and can each hold up to 20 individually-addressable
Yb\textsuperscript{+} ions, stretched out in chains (1D Coulomb crystals) along the null line of the two-dimensional
quadrupole RF driving field and along the mode of an optical cavity
(see Figure \ref{fig:trapfig1}). In addition, 24 DC electrodes on the microfabricated trap allow the shaping of quartic potentials along the cavity, as well as accurate micromotion compensation along the length of the array, which we measure spectroscopically using the optical cavity. We achieve deterministic control of ion distribution between the sites of the array, and control of ion-cavity coupling. Our measurements of fluorescence into the cavity indicate strong collective coupling with around 10 ions in a single array site. The demonstrated controls of micromotion, of distribution of ions between array sites, and of the cavity coupling are important for achieving high fidelity of interaction of the ion chains with the cavity photon bus \cite{Lamata2011a}, as well as high fidelity of phonon or ion exchange between the chains.

The optical cavity in our system can also be used to generate an intracavity
standing wave with a maximum intensity of 100 kW/cm\textsuperscript{2},
resulting in a deep all-optical periodic potential for the ions with
a periodicity of 185 nm, which is three orders of magnitude smaller than the
periodicity of the trap array. This makes our system attractive for QIP with ions in optical microtraps and for QSim with ions in periodic potentials.

\section{Experimental setup and trap fabrication}

The centrepiece of the experimental set-up is a planar-electrode
linear Paul trap placed between the mirrors of a 2.2-cm-long optical
cavity. The maximum finesse of the cavity is 12,500 and the TEM$_{00}$
mode waist is 38 $\mu$m, corresponding to a calculated antinode cooperativity
$\eta=0.23$ ($\eta$ is defined in section 4, equation \ref{eqn:eta}) for the 369 nm $^{2}$S$_{1/2}-^{2}$P$_{1/2}$ transition
in the Yb\textsuperscript{+} ion. The trapping region of the Paul
trap consists of the central 8 mm of the RF quadrupole nodal line, which is
overlapped with the mode of the cavity by careful alignment \cite{GrierThesis2011}. A split central electrode with a periodic structure, shown in Figure \ref{fig:trapfig1}c, allows
the long trap to be sectioned into 50 separate trapping sites along the
cavity mode, spaced by 160 $\mu$m, by applying a negative DC voltage to the inner periodic electrode and a positive DC voltage to the outer periodic electrodes. The ratio of outer electrode voltage to inner electrode voltage of -0.9 was chosen to cancel the displacement of the trap in the direction perpendicular to the chip surface.
Confinement in the plane perpendicular to the cavity axis is achieved
by applying an RF voltage of 127 V amplitude at 16.4 MHz to the two
long outer electrodes shown in green in Figure \ref{fig:trapfig1}d and
grounding the rest of the trap at RF frequencies. The resulting Paul trap resides
134 $\mu$m from the electrode surface, has trap frequencies $\omega_{x,y}$ = $2\pi\times$1.3 MHz, has a Mathieu parameter
$q$ = 0.22 and a trap depth of 84 meV along the weakest confinement direction away from the chip surface (pseudopotential
shown in Figure \ref{fig:trapfig1}e). In practice, the trap depth is usually increased to a few hundred meV by applying a DC quadrupolar field that provides additional confinement along the weak axis at the expense of deconfinement along the other axis. This is achieved by applying a negative DC voltage of a few volts to the RF electrodes. As result of this DC quadrupole field, the x- and y- trap frequencies differ by a few hundred kHz. The trap is equipped with 24
DC electrodes for finely shaping the potential along the linear dimension
of the trap, and for compensating stray electric fields in the RF
trapping (transverse) plane (Figure \ref{fig:trapfig1}b).

The trap was fabricated on a 500-$\mu$m-thick single-crystal
quartz wafer using a procedure similar to the one described in
\cite{Labaziewicz2008}. The wafer was mechanically cleaned using
Clorox 409 surfactant, followed by a 30 minute soak in a 3:1 solution
of H$_{2}$SO$_{4}$:H$_{2}$O$_{2}$ at room temperature and a 20 minute
soak in 4:1:1 H$_{2}$O:NH$_{4}$OH:H$_{2}$O$_{2}$ at 65$^{\circ}$C.
A 10 nm Ti adhesion layer was evaporated using an electron beam, followed by a
300-nm-thick Ag layer. The trap pattern was defined using 6-$\mu$m-thick AZ4620 positive photoresist exposed through a soft contact chrome
mask. The trapping electrodes were produced by electroplating a 1.7-$\mu$m-thick layer of Au onto the exposed Ag with 0.85 $\mu$A/cm$^{2}$
current density (single-polarity) using Transene TSG-250 sulfite Au-plating
solution in a stirred 49$^{\circ}$C bath. To electrically separate
the trapping electrodes, the AZ4630 photoresist was removed by acetone,
followed by a 20 second Ag etch in 1:1:4 H$_{2}$O$_{2}$:NH$_{4}$OH:H$_{2}$O
and a 5 second Ti etch in 1:4 HF:H$_{2}$O. The wafer was
then protected with 1 - 2 $\mu$m-thick NR9-3000P photoresist
and cut into individual 2.3 mm$\times$18 mm traps on a carbide diesaw. Given that our material processing steps were identical to the ones used in \cite{Labaziewicz2008}, we expect the heating rate of ions trapped 134 $\mu$m from the trap surface to be on the order of $2\pi\hbar$ $\times$ 1 MHz / ms.

\begin{figure}
\includegraphics[scale=0.35]{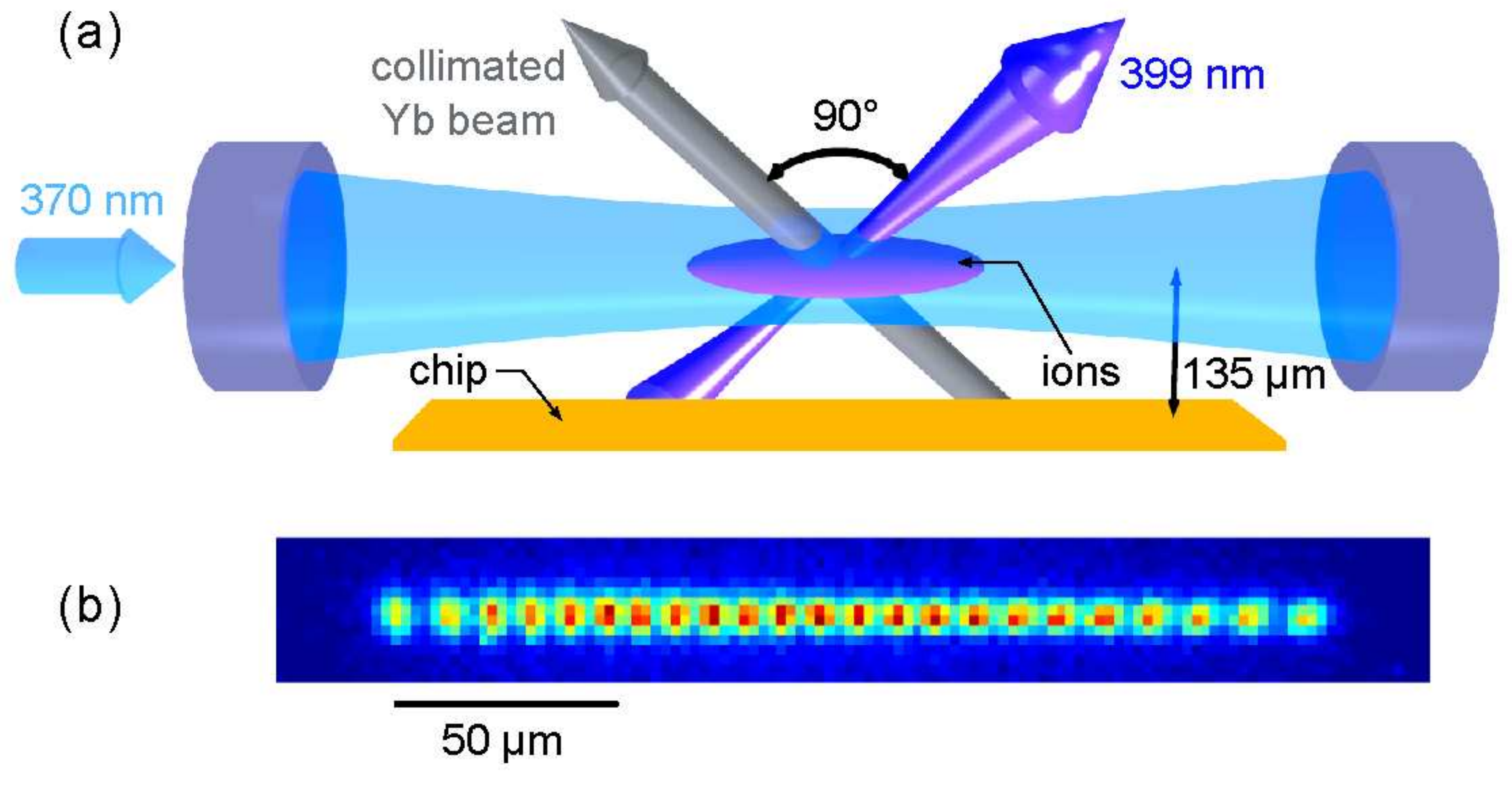}\caption{\label{fig:exp_setup}a) Trap loading configuration. 399 nm light
perpendicular to the thermal atom beam excites the isotope of choice
to the $^{1}$P$_{1}$ excited state, from which ionization to the continuum
proceeds via the cavity-enhanced 369 nm light. b) Isotopically pure
1D crystal of 23 ions of \textsuperscript{174}Yb\textsuperscript{+} in a harmonic potential. Ions of a different isotope would be off-resonant with the excitation light and would appear as dark gaps in the chain.}
\end{figure}

The cavity mirrors were coated by Advanced Thin Films with a dieletric
stack structure of SiO$_{2}$ and Ta$_{2}$O$_{5}$(top layer) with a
2:1 thickness ratio. Mirror transmission was quoted at T = 1.8$\times$10$^{-4}$
at 369 nm. Before the set-up was inserted into the vacuum chamber, the finesse was measured
to be $F_{0}=1.25\times10^{4}$. Under vacuum, the finesse has been
degrading steadily over time, with mirror loss (L) increasing linearly
at a rate of $6\times10^{-5}$/month ($F=\pi/(T+L)$). The experimental
results presented are for a finesse of $F_{1}=2400$. The finesse
may be restored by oxygen treatment \cite{Gangloff2012}
before implementing efficient quantum information protocols with ions,
but certain proof-of-principle experiments can be performed with the
current set-up.

Yb\textsuperscript{+} ions are produced by a two-step photo-ionization
\cite{Balzer2006, Cetina2007} of the effusive atom flux from a resistively heated
oven. The flux is collimated, and to avoid coating the trap, angled
away from the chip surface (see Figure \ref{fig:trapfig1}a). The ionization is accomplished by resonant
excitation on the 399 nm $^{1}$S$_{0}-^{1}$P$_{1}$ neutral Yb transition
perpendicular to the oven flux, combined with one-photon ionization
via 369 nm light built up in the resonator mode (see Figure \ref{fig:exp_setup}a).
The intra-cavity intensity of the 369 nm light can be continuously
controlled up to a maximum of 2 kW/cm$^{2}$, resulting in a loading
rate of $\thicksim$2 ions/second. Addressing the coldest direction
(perpendicular to the flux) of the atomic beam with the 399 nm light
minimizes Doppler broadening and resolves the different Yb isotopes
of interest, which are spaced by at least 250 MHz in frequency. This
allows us to achieve isotopic purity of our ion samples in excess of 90\% (see
Figure \ref{fig:exp_setup}b).\\

\begin{figure}
\includegraphics[scale=0.5]{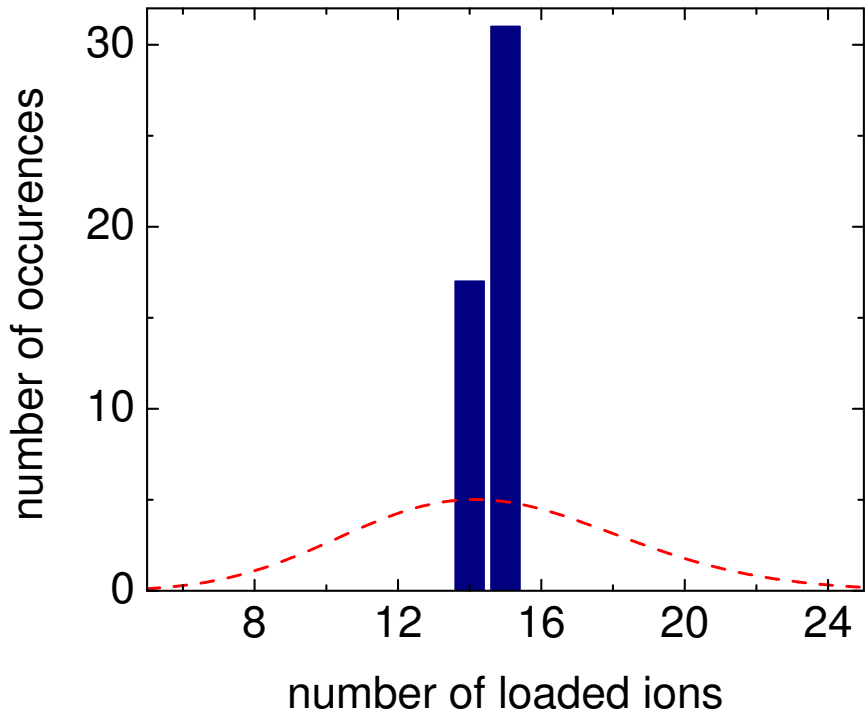}\caption{\label{fig:loading-stats}Typical ion number histogram for an array
site after splitting a longer chain repeatedly with the periodic potential.
The red dotted curve is the Poisson distribution with the corresponding
mean. The Fano factor at this site is 1.6\%. The Fano factor at any
site of the array is <10\%, limited by residual isotopic impurity.}
\end{figure}

\section{Deterministic splitting of ion crystals}

Our ability to control the loading
rate and the shape of the axial potential permits the loading of a controlled
number of ions into a long isotopically pure 1D crystal in the trapping region. By applying
a DC periodic potential, we can split this crystal between the individual
array sites with up to 20 ions per site. The crystal order of our
sample resulting from the strong ion-ion repulsion pins each ion with
respect to the applied periodic potential, causing the ion crystal to be split at defined positions and resulting in a deterministic loading of the trap array. By ramping the periodic potential up and down multiple
times and counting the number of ions loaded into each array site, we collect
statistics that show highly suppressed ion number fluctuations, as
shown on Figure \ref{fig:loading-stats}. This can be quantified using
the Fano factor, which is defined as the ratio of the ion number variance to the mean number of ions loaded in each
trap. This factor is unity for a Poisson process, which would be expected
for weak interactions. We typically observe Fano
factors less than 0.1. In addition, by shaping the overall axial potential
appropriately, we can control the mean number of ions loaded into
each trap. \\

\begin{figure}
\includegraphics[scale=0.4]{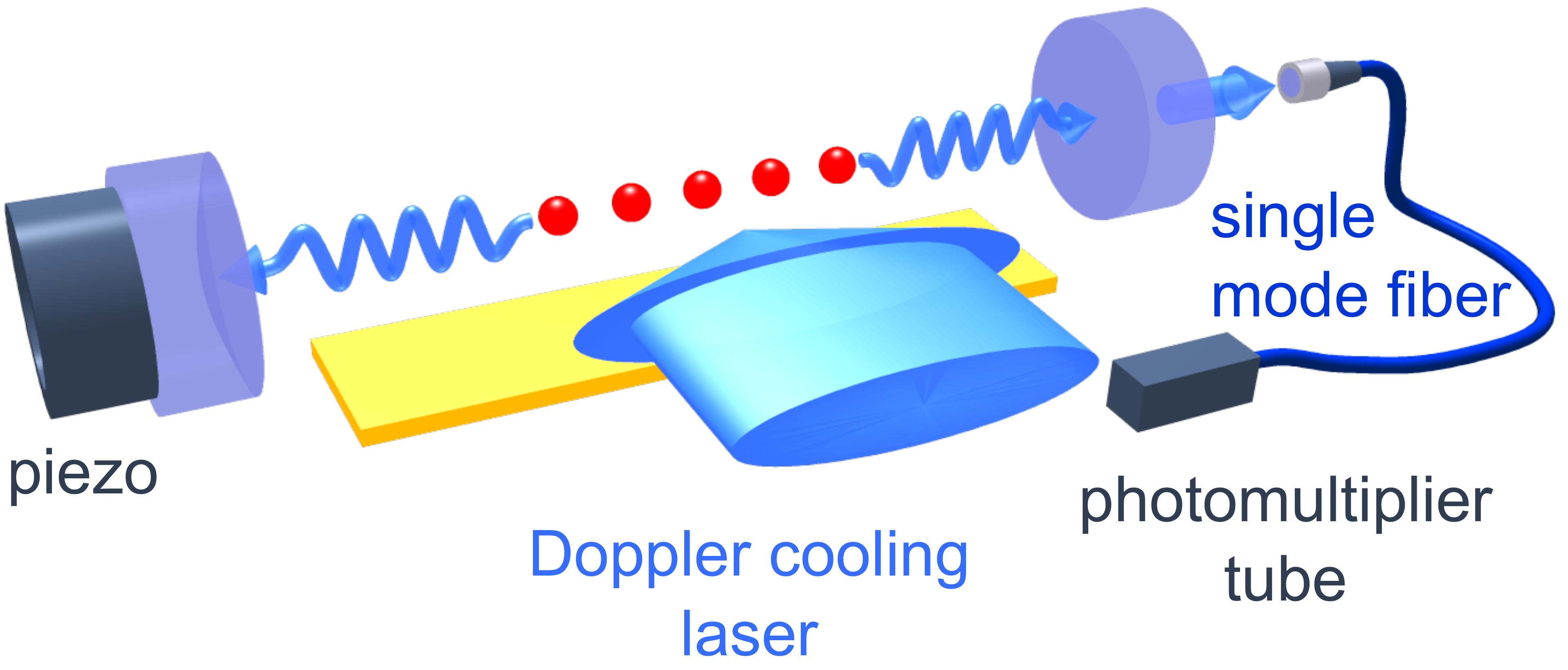}\caption{\label{fig:spectroscopy-set-up}Set-up for fluorescence spectroscopy
via the cavity. Doppler cooling light perpendicular to the cavity-trap
axes is scattered into the cavity, which acts as a tunable spectrometer.
The light exiting the cavity is coupled to a single-mode fiber and detected by a PMT. For results presented here, incident light is linearly polarized in the direction perpendicular to the cavity-trap axes, and collected photons are polarization-analyzed in the same direction.}
\end{figure}

\section{Ion-cavity coupling}

In free space, the probability
of dipole interaction between a photon and a two-level ion can be
quantified by the absorption probability, given by the ratio of the
resonant atomic photon scattering cross-section $\frac{3}{2\pi}\lambda_{0}^{2}$
to the mode area $\frac{1}{2}\pi w^{2}$ of a Gaussian laser beam,
where $\lambda_{0}=2\pi/k_{0}$ is the resonant wavelength and $w$
is the mode size \cite{HarukaAdvances2011}. In free space, this ratio
is substantially smaller than unity as diffraction limits the waist
size. One solution is an optical cavity, which enhances the interaction
probability by the number of round trips $F/\pi$ in the cavity (where
$F$ is the cavity finesse) and a factor of 4 at the antinode of the
standing wave. The result is the single particle antinode cooperativity
\cite{HarukaAdvances2011}
\begin{equation}
\label{eqn:eta}
\eta=\frac{24F}{\pi w^{2}k_{0}^{2}}=\frac{4g^{2}}{\kappa\Gamma}
\end{equation}
which is the figure of merit in cavity quantum electrodynamics, encapsulating the strength
of atom-cavity coupling $g$ (single photon Rabi frequency 2$g$) relative
to the cavity decay rate constant $\kappa$ and the atomic spontaneous
emission rate constant $\Gamma$. (Note that the second expression
can be generalized to any transition in a multi-level ion by using
the appropriate $g$ and $\Gamma$, while the first expression is only applicable to a 2-level system.) A coherent ion-photon interface
necessitates the strong coupling regime $\eta\gtrsim1$, which can
be achieved with a high finesse $F$ (by increasing the mirror reflectivity),
or by using micro-cavities with a small waist $w$. Alternatively, by coupling a cavity to an ensemble of $N$ ions, the collective cooperativity
is enhanced by the factor $N$ and the strong collective coupling regime is achieved for $N\eta\gtrsim1$.

\begin{figure}
\includegraphics[scale=0.5]{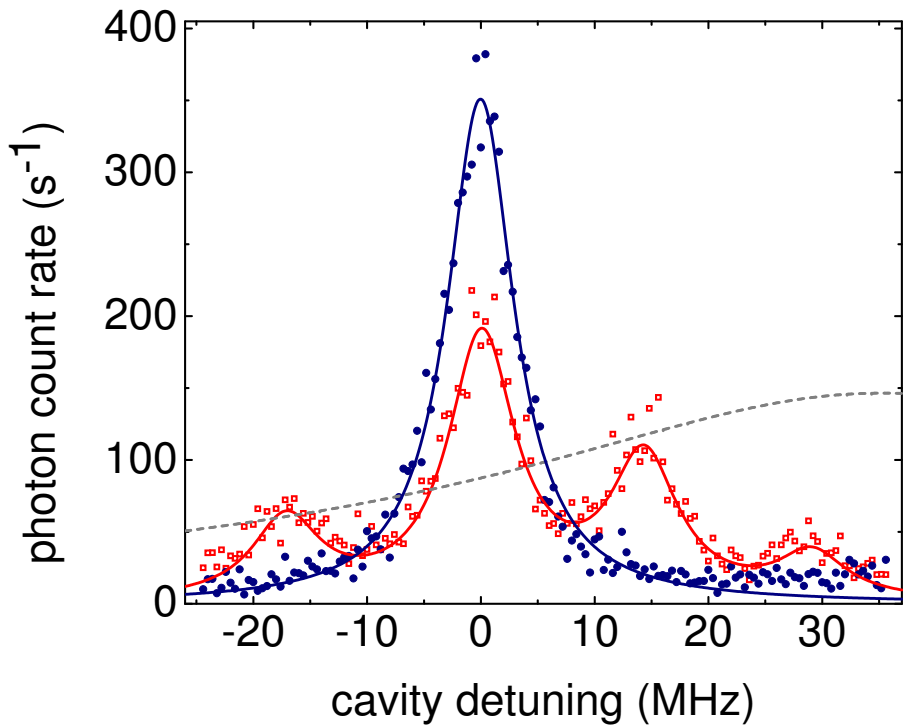}\caption{\label{fig:cav-fluor-spectra}Single-ion fluorescence spectra collected
by the cavity for a well-compensated ion (blue) and an ion decompensated
(red) in the direction of the Doppler cooling beam by a DC electric
field of 65 V/m. The 1st-order micromotion sidebands show up clearly
and a 2nd-order sideband is also visible on one side. The asymmetry
in the spectrum is due to higher scattering rate on the side closer
to atomic resonance; the grey dotted curve shows the calculated
dependence of the photon scattering rate on the detuning.}
\end{figure}

In order to quantify the coupling between the trapped ions and the
cavity mode in our system, we use the set-up shown on Figure \ref{fig:spectroscopy-set-up}. $^{174}$Yb$^{+}$ ions held in the planar trap are Doppler-cooled on the $^{2}$S$_{1/2}$-$^{2}$P$_{1/2}$ transition ($^{2}$P$_{1/2}$ state natural linewidth $\Gamma$= $2\pi\cdot19.9$ MHz) using a linearly-polarized laser beam perpendicular to the cavity-trap axis with polarization also perpendicular to the cavity-trap axis. The intensity of the cooling light is I = 500 mW/cm$^{2}$, corresponding to the resonant saturation parameter $s_{0} = I/I_{\textnormal{\scriptsize sat}} = 10$, where $I_{\textnormal{\scriptsize sat}}$ = 51.5 mW/cm$^{2}$ is the saturation intensity of the transition. The ion fluorescence photons scattered into the cavity mode are polarization-analyzed, coupled into a single-mode fiber and counted using a photomultiplier tube (PMT). When the cavity is resonant with the cooling light and detuned from the atomic resonance by $\delta$ = 2.5 $\Gamma$, and when only photons with the same linear polarization as the incident light are collected, we observe 350 counts/second from a single ion in a weak axial potential (peak value in Figure \ref{fig:cav-fluor-spectra}). The ion in this case is not localized with respect to the cavity standing wave. The collected photon flux is consistent with the expected cooperativity of $\eta$ = 0.044 for a finesse of $F_{1}=2400$, as explained below.

The total scattering rate into free space for the given saturation
and detuning is $\Gamma_{sc}=\frac{s}{1+s}\frac{\Gamma}{2}=1.7\cdot10^{7}$
$s^{-1}$, of which a fraction $1/(1+s)$ = 0.72 is coherent, where $s = s_{0}/(1+(\frac{2\delta}{\Gamma})^2)$ is the saturation parameter \cite{API}.
Of these, a fraction $\frac{1}{2}\eta\cdot\kappa/(\kappa+\Delta\omega_{\textnormal{\scriptsize laser}})$
is collected by the cavity, where the antinode cooperativity $\eta$
is multiplied by 1/2 for averaging over the cavity standing wave,
and by the spectral overlap of the cavity and the laser. The cavity
linewidth is $\kappa$ = $2\pi\times2.7$ MHz and the laser linewidth
$\Delta\omega_{\textnormal{\scriptsize laser}}$ is $2\pi\times4.9$ MHz. The incoherently
scattered photons are broader in frequency than the cavity linewidth
and are not collected efficiently. Collecting light of the same linear
polarization as the excitation beam results in a reduced dipole matrix
element squared of 1/3 for the $J=\frac{1}{2}\rightarrow J=\frac{1}{2}$
transition. In addition 1/2 of the photons are lost through the other
port of the cavity and only $T/(T+L)$ = 13\% of the photons are coupled
out of the cavity due to losses on the mirrors. Finally, after including
cavity-to-fiber mode matching of 0.9, transmission through all the
optics of 0.7 and PMT quantum efficiency of 0.28, the expected count
rate of $\thicksim$375 counts/second,  matches the observed count rate of $\thicksim$350 counts/second within 10\%.

For purely coherent processes within the cavity, such as the transfer
of quantum information between ion chains via the cavity mode, of
the stated factors degrading our photon collection efficiency, only the reduced dipole matrix element squared and the cavity standing wave
averaging are relevant. By localizing the ion with respect to the
cavity standing wave, the full antinode cooperativity $\eta$ may be recovered, modulo the reduced
dipole matrix element. An additional advantage of localizing the ions to better than the mode wavelength (Lamb-Dicke
regime) is to reduce motional decoherence. This approach has been
used to extend light storage times in atomic ensembles from microseconds
\cite{Tanji2009} to a hundred milliseconds by localizing them in optical
lattices with the added challenge of careful AC Stark shift compensation \cite{Kuzmich08,Radnaev2010}.
The longest light storage time of 240 ms was achieved in a Mott insulator,
limited by tunneling and lattice heating \cite{Schnorrberger2009},
which, together with AC Stark shifts, are absent or negligible in
a Paul ion trap.

\begin{figure}
\includegraphics[scale=0.5]{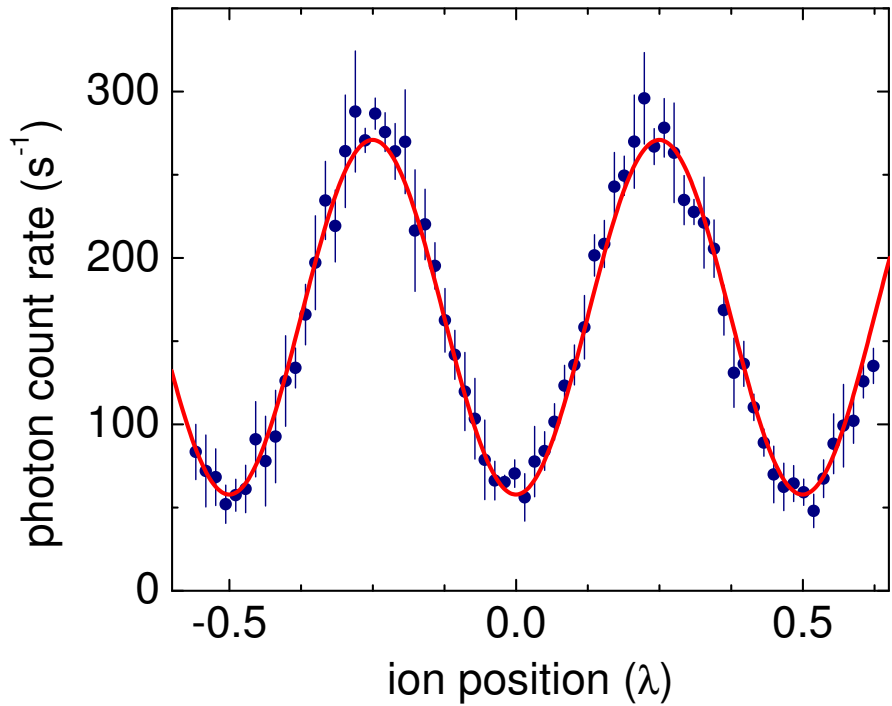}\caption{\label{fig:fluor_modulation}Single-ion fluorescence collected by
the cavity resonant with the excitation laser as a localized ion is transported along the
cavity mode. The observed mode visibility is 65\%, corresponding
to an ion temperature of 1.6 $T_{\textnormal{\scriptsize Doppler}}$ and a thermal RMS spread of the ionic wavepacket of 27 nm.}
\end{figure} 

We demonstrate precise positioning with respect to the cavity
standing wave by confining the ion in the strong axial potential
of the periodic array with an axial vibration frequency of $\omega_{\textnormal{\scriptsize ax}}=2\pi\times$1.14
MHz, and moving it along the cavity axis. We observe a periodic variation
of fluorescence scattered into the cavity (shown in Figure \ref{fig:fluor_modulation}),
which corresponds to the spatial modulation of ion-cavity coupling
between the node and antinode of the cavity mode (a similar technique
was used in \cite{Guthohrlein2001}). For the ion in Figure \ref{fig:fluor_modulation},
the fringe visibility is 65\%, corresponding to a temperature of 1.6$T_{\textnormal{\scriptsize Doppler}}$
(where $T_{\textnormal{\scriptsize Doppler}}=\frac{1}{2}\hbar\Gamma/k_{B}$ is the Doppler limited temperature)
and an ion localized to 27 nm (wavefunction RMS spread) and a recovery of antinode $\eta$ to 80\%. These numbers
could be further improved using sideband cooling techniques.

With the observed value $\eta$ = 0.044, corresponding to a cavity
finesse of $F_{1}=2400$, the effective cooperativity parameter approaches
unity for an ensemble of about 20 ions. At the initial, undegraded cavity finesse
of $F_{0}$ = 12,500, this is reduced to 5 ions per array site.\\

\section{Motional spectroscopy}

The spectrum of light scattered into the cavity mode by the ions can
be measured by scanning the cavity length at a fixed ion excitation
laser frequency (see Figure 4). The resolution is limited only by
the convolved linewidths of the drive laser $\Delta\omega_{\textnormal{\scriptsize laser}}$
and of the cavity $\kappa$, which combine to give $\Delta\omega_{\textnormal{\scriptsize res}}=2\pi\times7.5$
MHz. This is sufficiently narrow to resolve the micromotion sidebands
resulting from the RF trap drive at 16 MHz and to accurately minimize
them by compensating the position of the ion relative to the RF null
using the DC electrodes (Figure \ref{fig:cav-fluor-spectra}). This
step is important in minimizing motional heating of ions. In addition,
it allows us to verify that there is no parasitic micromotion along
the cavity axis, which would degrade the fidelity of interaction between
ions and cavity photons by spreading the ion spectrum into the micromotion
sidebands. This method of micromotion compensation is an alternative
to more standard techniques \cite{Berkeland1998} and gives a direct
measure of the sideband strength relative to the carrier. (Note that
a similar method was used in \cite{Stute2012b}.) \\

\section{Conclusion}

We have built and characterized a novel hybrid quantum
system where ion-based quantum information processing can be interfaced
with a phonon bus between microtraps or a cavity-based photon bus, and where light forces could be used
to construct 1D Hamiltonians of particles with strong long-range interactions
in a periodic potential. This system presents a rich Hilbert
space spanned by the internal atomic, phononic and photonic degrees
of freedom, where competing interactions could be studied in addition
to digital quantum computation.

We acknowledge resources provided by the research group of Karl Berggren
in the ion trap fabrication process. This work was supported by the
Army Research Office (ARO) and the National Science Foundation (NSF).
A.B. and D.G. gratefully acknowledge support by the National Science
and Engineering Research Council of Canada (NSERC) and L.K. gratefully
acknowledges support by the Alexander von Humboldt Foundation. K.B. gratefully acknowledges support from the National Science Foundation (NSF) through the Graduate Research Fellowship (0645960) and Interdisciplinary Quantum Information Science and Engineering (0801525) programs.

\section*{References}

\bibliographystyle{iopart-num}
%\bibliography{\string"C:/Documents and Settings/alexeibyl/My Documents/Bibliography/library\string"}
\bibliography{library_final}

\end{document}